\documentstyle[aps,prl,multicol,epsf]{revtex}

\begin{document} 

\title{Random Dirac Fermions and Non-Hermitian Quantum Mechanics}

\author{Christopher Mudry and B. D. Simons\cite{Present}}
\address{Lyman Laboratory of Physics, Harvard University, MA 02138, USA}

\author{Alexander Altland}
\address{Institut f\"ur Theoretische Physik, Universit\"at zu K\"oln,
Z\"ulpicher Strasse 77, 50937 K\"oln, Germany}

\date{December 8, 1997}
\maketitle 
 
\begin{abstract}
We study the influence of a strong imaginary vector potential on the quantum 
mechanics of particles confined to a two-dimensional plane and propagating in 
a random impurity potential. We show that the wavefunctions of the 
non-Hermitian operator can be obtained as the solution to a two-dimensional
Dirac equation in the presence of a random gauge field. Consequences for the
localization properties and the critical nature of the states are discussed.
\end{abstract}


\begin{multicols}{2}  

The field of non-Hermitian quantum mechanics
has attracted great interest 
recently both in its connection to anomalous diffusion in random 
media~\cite{Isichenko92}, as well as to the statistical mechanics of flux 
lines in superconductors~\cite{Hatano96}.
At the same time, motivated in part
by the connection to properties of the integer quantum Hall transition
and gapless superconductors detailed investigations have been made 
into the critical properties of Dirac Fermions coupled to random gauge fields
\cite{Fisher85,Ludwig94,Nersesyan94,Bernard95,Mudry96,Chamon96,Kogan96,Caux97,Stephanov96}.
The aim of this letter is to identify a connection between these problems 
which explains some of the unusual phenomena recently reported in the 
behavior of two-dimensional non-Hermitian random operators~\cite{Hatano96}. 
This correspondence is related to a {\it chiral symmetry} of  
effective Hamiltonians commonly used in the analysis of problems in
non-Hermitian quantum mechanics.

The two-dimensional Hamiltonian we consider describes a particle propagating 
in a random scalar impurity potential, $V$ and subject to a uniform imaginary 
vector potential, $i{\bf h}$
\begin{eqnarray}
\hat{H}={1\over 2m}\left(\hat{\bf p}+i{\bf h}\right)^2+V({\bf r}).
\label{hamil}
\end{eqnarray}
The scalar potential, $V$ is assumed to be real and drawn from some random 
distribution, $P[V]$ which, for now, is left unspecified. Related problems 
have been recorded in a variety of physical situations ranging from the study 
of reaction diffusion phenomena in biological systems~\cite{Shnerb97}, 
to advective diffusion in random media~\cite{Isichenko92,Fisher84,Chalker97}, 
and the study of fluctuating vortex lines in superconductors with columnar 
defects~\cite{Hatano96}.

Remarkably, in contrast to properties of the Hermitian operator (i.e. one in
which the vector potential is real), numerical studies~\cite{Hatano96} 
suggest localization properties of $\hat{H}$ depend sensitively on the 
relative strength of ${\bf h}$. While, in dimensions $d\le 2$, {\it all}
wavefunctions of the Hermitian operator are believed to be 
localized~\cite{Lee85}, numerical evidence suggests that application of a 
sufficiently strong vector potential, ${\bf h}$ induces a delocalization 
transition of states of the non-Hermitian operator. In contrast to the 
situation in $1d$~\cite{Hatano96,1dcase}, 
the mechanism and stability of the 
delocalization transition is not yet well understood.

Recent analytical studies have focussed on the spectral properties of the 
non-Hermitian Hamiltonian~\cite{Efetov97}. Treating the vector potential as a 
weak perturbation of the random Hamiltonian, impurity averaged properties of 
the Green function have been cast in the form of a functional field integral 
involving a supersymmetric non-linear $\sigma$-model. While capturing {\em 
universal} features of the complex spectrum of $\hat{H}$, evidence for 
delocalization of states was not sought. By contrast, in the present approach, 
we will impose a strong imaginary vector potential, and treat the random 
potential as a perturbation
(a regime explored in 1$d$ by Feinberg and Zee~\cite{1dcase}). 
By doing so we will reveal an explicit connection 
between the Hamiltonian in Eq.~(\ref{hamil}) and the problem of Dirac fermions 
propagating in a random gauge field. In particular, we will find a regime
of weak disorder in which it is possible to construct explicitly 
eigenfunctions of the non-Hermitian Hamiltonian $\hat H$ 
for {\it individual} realizations of the disorder. 

In the absence of the impurity potential, the eigenfunctions of $\hat{H}$
are plane waves with complex eigenvalues, 
$z_0({\bf p})=({\bf p}^2-{\bf h}^2)/2m+
i{\bf p}\cdot{\bf h}/m$. 
Since the Hamiltonian is real ($\hat{H}=\hat{H}^*$), eigenvalues occur in 
complex conjugate pairs, a property maintained in the presence of the random
potential. In the infinite system, the spectrum forms a dense support  
occupying the region of the complex plane 
$|{\rm Im}\ z|\leq\sqrt{2m{\rm Re}\ z+|{\bf h}|^2}\ |{\bf h}|/m$.
The two-dimensional density of states (DoS) takes the form 
\begin{eqnarray}
\nu_0(z)={\nu_0\over \pi} {m\over\left[(2m{\rm Re}\ z+
{\bf h}^2) {\bf h}^2-m^2 
({\rm Im}\ z)^2\right]^{1/2}},
\end{eqnarray}
where $\nu_0$ denotes the constant DoS of the Hermitian Hamiltonian.

Concerned with impurity averaged spectral properties of the Hamiltonian, we 
begin by defining the Green function $\hat{G}(z)=(z-\hat{H})^{-1}$
where $z$ denotes the complex argument. Using 
$\nu(z)\equiv (1/\pi\Omega){\rm tr}\ \partial_{z^*}G(z)$, where $\Omega$
represents the volume, $G(z)$ is shown to be non-analytic everywhere 
the DoS $\nu(z)$ is non-vanishing \cite{Chalker97,Sommers88}. 

To properly account for non-analytic properties of the 
impurity averaged Green function, previous 
studies~\cite{Chalker97,Efetov97,Sommers88} have emphasized the 
need to express the Green function through a Hamiltonian which is explicitly 
Hermitian. This is achieved by constructing a matrix Hamiltonian with the 
$2\times 2$ block structure 
\begin{eqnarray}
\hat{\cal H}=\left(\matrix{0&\hat{H}-z\cr \hat{H}^\dagger -z^*&0\cr}\right).
\label{cal H}
\end{eqnarray}
In this representation, the Green function of the non-Hermitian operator is
expressed as the off-diagonal element of the matrix Green 
function, 
$\hat{G}(z)=\lim_{\eta\to 0}{\hat{\cal G}}_{21}(z)$,
where, defining $\eta=0^+$, $\hat{\cal G}=(i\eta-\hat{\cal H})^{-1}$.
Zero energy eigenstates of the matrix Hamiltonian $\hat{\cal H}$
yield eigenstates of the non-Hermitian Hamiltonian $\hat H$.

Defining Pauli matrices $\vec{\sigma}$ ($\sigma_0=\openone$) which operate in
the $2\times 2$ space, the {\em chiral} symmetry of the matrix Hamiltonian, 
$\hat{\cal H}=-\sigma_3\hat{\cal H}\sigma_3$ implies that 
eigenvalues, $\epsilon_i$ 
of $\hat{\cal H}$ appear in pairs of opposite sign. Moreover, any such 
pair of eigenstates obeys $|\psi_{+\epsilon_i}\rangle=\sigma_3 
|\psi_{-\epsilon_i}\rangle$. 

\begin{figure}[hbt]
\centerline{\epsfxsize=3in\epsfbox{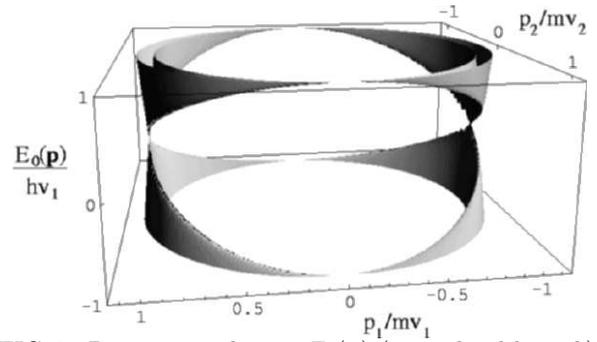}}
\caption{Dispersion relation, $E_0({\bf p})$ (normalized by $v_1 h$) in
the vicinity of zero energy shown as a function of ${\bf p}$ with 
${\bf h}=h\hat{\bf e}_2$, ${\rm Im}\ z=0$ and $mv_1/2h=10$. Note that the 
minimum scattering amplitude required to smear out the points of degeneracy
is given by $\hbar/\tau\sim hv_1$.}
\label{fig:support}
\end{figure}

The spectrum of $\hat{\cal H}$ depends sensitively on the strength of the 
vector potential, ${\bf h}$. In the absence of the random potential, the 
dispersion relation takes the form,
\begin{eqnarray}
E_0({\bf p})=\pm|z_0({\bf p})-z|,
\label{spec}
\end{eqnarray}
invariant under reflection about the axis parallel to ${\bf h}$. Thus,
in contrast to a Hamiltonian involving a real vector potential, where
the spectrum is described by a simple shift of the Fermi sphere, the
continuous degeneracy of the zero eigenvalues (the poles of the Green
function) is lifted. Instead, setting ${\bf h}\equiv h{\bf e}_2$, zero
energy states exist at only {\em two} discrete points 
${\bf p}_0^{(a)}$, $a=1$, $2$~\cite{foot2} (see Fig.~\ref{fig:support}).

If the impurity potential is strong (i.e. $\ell\ll \hbar/h$, where $\ell$ 
denotes the transport mean free path associated with the random impurity 
potential, $V$), unperturbed states of the clean system are strongly mixed by 
the disorder (see Fig.~\ref{fig:support}). In this limit, the pole structure of
the impurity averaged Green function, $\hat{\cal G}$ is smeared out.  
Correspondingly, statistical properties of $\hat{\cal H}$ are largely 
insensitive to the zero eigenvalues of $E_0({\bf p})$. In this limit, one 
can expect the transport properties of the non-Hermitian operator to reflect 
those of the Hermitian counterpart. Conversely, if the impurity potential is 
weak ($\ell\gg \hbar/h$) the pole structure of the average Green function 
is dominated by the nature of the spectrum in the vicinity of the zeros, 
${\bf p}_0^{(a)}$.

\narrowtext

Focusing on the limit of weak disorder 
\begin{eqnarray}
{\rm Re}\ z\gg {h^2\over 2m}\gg {\hbar\over \tau},{\rm Im}\ z,
\label{regime of interest}
\end{eqnarray}
where $\tau$ represents the corresponding mean free scattering time of the
random potential, a linearisation of the spectrum in the vicinity of the two 
zero eigenvalues separates the spectrum into two branches. Treating the random 
potential and ${\rm Im}\ z$ as weak perturbations, and performing a gradient 
expansion in
\begin{eqnarray}
\hat{\bf p}^{(a)}=
{\bf p}_0^{(a)}+(-1)^a\hat p^{\ }_1{\bf e}^{\ }_1+\hat p^{\ }_2{\bf e}^{\ }_2,
\label{momentumdefs}
\end{eqnarray}
where ${\bf p}_0^{(a)}=(-1)^{a-1} mv_1{\bf e}^{\ }_1$, $mv_1=(2m{\rm Re}\ z+
h^2)^{1/2}$, and $mv_2=h$, the low energy, long-wavelength expansion of the 
(unperturbed)
Hamiltonian around the Fermi points ${\bf p}_0^{(a)}$ generates the 
{\em anisotropic} Dirac operator
\begin{eqnarray}
\hat{\cal H}^{(0)}_D=-
\tau_0\otimes 
\left[
\sigma_1v_1(\hat{\bf p}\cdot{\bf e}_1)+
\sigma_2v_2(\hat{\bf p}\cdot{\bf e}_2)\right].
\label{diracham0}
\end{eqnarray}
Here we have introduced an additional set of Pauli matrices,
$\vec{\tau}$ ($\tau_0=\openone$), that index the block structure
associated with the reflection symmetry. The existence of two
degenerate zero energy eigenstates
of $\hat{\cal H}^{(0)}_D$ conspires with the
exact chiral symmetry to yield an anisotropic Lorentz symmetry.

Being generally non-symmetric under reflection, matrix elements of the random 
impurity potential violate the reflection symmetry. Accounting for matrix 
elements which scatter across the Fermi surface (i.e. between the Dirac 
points) as well as within each subspace, the general Hamiltonian takes 
the form
\begin{eqnarray}
\hat{\cal H}^{\ }_D=
\hat{\cal H}^{(0)}_D+
\sum_{\nu=0}^3 \tau_\nu\otimes\sigma_1\ V_\nu +
\tau_0\otimes\sigma_2\ {\rm Im}\ z,
\label{diracham}
\end{eqnarray}
where the impurity potentials, $V_\nu({\bf r})$ are real random functions
with Fourier components 
\begin{eqnarray}
\sum\limits_\nu
\tilde V_\nu({\bf q})\tau_\nu\!=\!
&&
\pmatrix
{
\tilde V  (      -q_1,q_2)&
\tilde V  (q_1-2mv_1,q_2)\cr
\tilde V^*(q_1-2mv_1,q_2)&
\tilde V  (      +q_1,q_2)\cr
}.
\nonumber
\end{eqnarray}
Equations~(\ref{diracham0},\ref{diracham}) 
represent an important intermediate result of this Letter: 
Firstly, the low energy sector of
the original Hermitian Hamiltonian (\ref{cal H}) has been described
in terms of the stochastic Dirac Hamiltonian, $\hat{\cal H}_D$. 
Secondly, $\hat{\cal H}_D$ possesses the chiral symmetry 
$\sigma_3\hat{\cal H}_D\sigma_3=-\hat{\cal H}_D$, 
a direct consequence of the chirality of the auxiliary operator
(\ref{cal H}). The significance of the second observation lies in the
fact that the behavior of chiral Dirac Hamiltonians can be analyzed {\it
for individual realizations of the disorder}: As we shall see, it is 
possible to construct explicit solutions for the
zero energy eigenfunction $|\Psi_D\rangle$ of $\hat{\cal H}_D$. From these the
eigenfunctions $|\Psi_{z^{\ }}\rangle$ ($|\Psi_{z^*}\rangle$)
of the original non-Hermitian operator $\hat H$ ($\hat H^{\dagger}$)
can then be obtained as
\begin{eqnarray}
&&
|\Psi_{z}\rangle\approx 
\left(\begin{array}{cc}
0&\\
&1\end{array}\right)\!
\hat\Pi^{\dagger}
|\Psi_D\rangle,\
|\Psi_{z^*   }   \rangle\approx 
\left(\begin{array}{cc}
1&\\
&0\end{array}\right)\!
\hat\Pi^{\dag}
|\Psi_D\rangle,
\label{eigenstate of H dagger}
\end{eqnarray}
for an {\it arbitrary} eigenvalue $z$ provided Eq. (\ref{regime of interest}) 
holds. Here the matrix structure refers to the $\sigma$--space and the
operator $\hat\Pi$ accounts for the fact that in order to obtain
eigenfunctions of the full problem, the eigenstates of the low energy
expansion~(\ref{diracham}) have to be `boosted' to the Fermi points
${\bf p}^{(a)}_0, a=1,2$, respectively\cite{foot3}.

We next turn to the explicit construction of the wavefunction 
$\Psi_D({\bf r})$. To this end we first remove the anisotropy of the
Hamiltonian~(\ref{diracham0}) by rescaling the coordinates according
to $r_\mu=(v_\mu/v)x_\mu$, $v=$$\sqrt{v_1 v_2}$. 
As a result $\hat{\cal H}_D$ takes the canonical form 
\begin{eqnarray*}
\hat{\cal H}_D=
\sum_{\mu=1}^2\sigma_\mu\otimes
\left(i\tau_0\partial_\mu+{\cal A}_\mu\right),
\end{eqnarray*}
where $\hbar=v=1$ and the two components ($\mu=1,2$),
${\cal A}_\mu=A_\mu+B_\mu+C_\mu$
are given by
\begin{eqnarray*}
A_\mu=\delta_{\mu1}\tau_0 V_0,\quad 
B_\mu=\delta_{\mu1}\sum_{\nu=1}^3\tau_\nu V_\nu,\quad
C_\mu=\delta_{\mu2}\tau_0{\rm Im}\ z.
\end{eqnarray*}
The disorder in $\hat{\cal H}_D$ appears in the form of a minimal coupling to 
a generally non-abelian vector potential (i.e. $B_\mu\ne 0$). It is thus 
natural to seek a gauge transformation that removes the stochastic components 
of the Hamiltonian. Indeed, although the potentials, $A_\mu$ and $B_\mu$ are 
not in general of pure gauge type, the non-gauge components can be accounted 
for by extending the concept of gauge transformations so as to include 
`axial' transformations. 

Focusing on the abelian sector first, we decompose $A_\mu$ into a transverse 
(axial gauge) and a longitudinal (pure gauge) component: 
$A_\mu=\epsilon_{\mu \nu}\partial_\nu \chi_\bot+\partial_\mu \chi_\|$,
respectively \cite{foot4}. It is then straightforward to verify that 
\begin{equation}
\Psi_D({\bf x})=
e^{i\sigma_0 x_2{\rm Im}  z }
e^{i\sigma_0\chi_\| ({\bf x})}
e^{ \sigma_3 \chi_\bot({\bf x})}
\pmatrix{\Theta_+\cr\Theta_-}
\label{solutionA}
\end{equation}
represents a solution of $\hat{\cal H}_D \Psi_D=0$ for 
$B_\mu=0$~\cite{Aharonov79}. 
Here, $\Theta_\pm \in {\bf C}$ and the
gauge transformation $\exp(i\sigma_0x_2{\rm Im} z)$ has been used to 
dispose of the small imaginary component of the eigenvalue $z$ in the 
non-Hermitian problem. This contribution can be absorbed into the `boost' 
component of $\hat\Pi$~(\ref{eigenstate of H dagger})~\cite{foottop}. 

The treatment of the non-abelian components $B_\mu$ is conceptually similar 
but -- due to their non-commutativity in $\tau$-space -- technically more 
involved. Referring to Ref.~\cite{Roskies81} for details, we merely state 
that in the presence of finite $B_\mu$, Eq.~(\ref{solutionA}) generalizes to 
\begin{eqnarray}
\Psi_D({\bf x})&&=
e^{i\sigma_0 x_2{\rm Im}  z }\
e^{i\sigma_0\chi_{\|} ({\bf x})} U({\bf x})\
e^{ \sigma_3\chi_\bot({\bf x})} 
e^{ \sigma_3 \vec{\xi}({\bf x})\cdot\vec{\tau}}
\nonumber\\
&&\times
\pmatrix{f(x_1+ix_2)\ \Theta_+\cr g(x_1-ix_2)\ \Theta_-}.
\label{general solution}
\end{eqnarray}
Here, $U({\bf x})\in SU(2)$ plays the role of the abelian gauge transformation 
$\exp[{i\sigma_0\chi_\|({\bf x})}]$, whereas the vector field $\vec{\xi}$ 
plays the role of the abelian axial component $\chi_\bot$ and obeys an 
analogous but more complicated equation~\cite{Roskies81}. Finally, $f$ and 
$g$ represent analytic functions which are fixed by the boundary conditions 
(see below).

Equation~(\ref{general solution}) represents a non-perturbative solution 
to the zero energy Dirac equation for {\em any} given realization of the
disorder. It, therefore, allows the construction of the eigenfunctions
of the non-Hermitian Hamiltonian according to 
Eq.~(\ref{eigenstate of H dagger}). The most important properties of 
$\Psi_D$ are a) that the generalized gauge factors depend in a non-local way 
on the spatial distribution of the disorder potential and b) that the axial 
gauge factors lead to an exponential disorder dependent amplification of the 
{\it modulus} of the wavefunction, 
$
\Psi^{\dagger}_D\Psi^{\ }_D\equiv|\Psi^{\ }_D|^2=
\Theta^{\dagger} e^{2\sigma_3[\chi_\bot+\vec{\xi}\cdot\vec{\tau}]}\Theta
$. 

What can be said about the asymptotic behavior of the wavefunction 
$\Psi_D$, and, in particular, about its localization properties? To address 
this question we consider the impurity averaged two-point correlation 
function, 
$
C({\bf r}_1-{\bf r}_2)\equiv
\overline{
|\Psi_D({\bf r}_1)|^{2q_1}
|\Psi_D({\bf r}_2)|^{2q_2} }
$. 
From the structure of the solution we infer 
that $C=F_A\times F_B$ factorizes into an abelian and a 
non-abelian component both of which can be straightforwardly extracted from 
Eq.~(\ref{general solution}). To say more, it is necessary to specify both 
the form of the random potential distribution function, and the topology of 
the system. 
Here we focus on the thermodynamic limit 
with Gaussian white-noise distributed disorder, $P[V]$ of 
uniform variance. Under these circumstances, taking $f=g=1$ in 
Eq.~(\ref{general solution}) represents the only admissible 
choice~\cite{foot7}.

Defining $g_A=(2\times 2\pi\nu\tau)^{-1}$ as the variance of the 
coarse-grained Gaussian distribution for $A_\mu$
(i.e. 
$\overline{\chi_\perp({\bf x})\chi_\perp({\bf y})}
\propto-g_A\ln|{\bf x}-{\bf y}|$), 
and focusing, for 
simplicity, on the case $q_1=q_2=1$, the impurity average yields a strongly 
anisotropic ($v_1\gg v_2$)
{\it algebraic} decay,
\begin{eqnarray}
F_A({\bf r})\propto |(v_1 r_1)^2+(v_2 r_2)^2|^{-{2g_A\over\pi}}, 
\label{Abelian average}
\end{eqnarray}
implying that eigenstates of the non-Hermitean operator
$\hat H$ are {\it critical}~\cite{Mudry96,Chamon96}. Moreover, since the 
algebraic decay of $F_A$ is not influenced by ${\rm Im}\ z$, we 
infer that 
the critical nature of the 
wavefunctions is also insensitive to ${\rm Im}\ z$.

As for the non-abelian sector, the non-trivial relationship between the 
fields $B_\mu$ and the effective `gauge' field, $\vec{\xi}$ makes the 
calculation of the correlation function, $F_B$ more involved (c.f. 
Refs.~\cite{Nersesyan94,Bernard95,Mudry96,Caux97}). However, although at
present no rigorous statements on the long distance behavior 
of the correlation function $F_B$ can be made in general, insight can be 
drawn from the following facts: a) The theory possesses a strong coupling 
fixed point described by the Wess-Zumino-Novikov-Witten theory 
$SU_{-4}(2)$~\cite{Nersesyan94,Mudry96} (corresponding to an infinite 
strength of the non-abelian components of the disorder potential), and b) 
in this limit~\cite{Mudry96,Caux97}, the correlation function,
\begin{eqnarray}
F_B({\bf r})\propto |(v_1 r_1)^2+(v_2 r_2)^2|^{-1/2}
\label{Non Abelian average}
\end{eqnarray}
is again algebraic. Note that the scaling exponent in
Eq. (\ref{Non Abelian average}) 
is fixed solely by the number of nodes (two) in Eq. (4).
Therefore, given that the wavefunctions are critical in 
the strongly disordered limit, it seems highly plausible that they remain 
critical in general.

In this Letter, we have studied the spectral properties of the two-dimensional 
random Schr\"odinger operator in the presence of a uniform imaginary vector 
potential, ${\bf h}$. Mapping it to a Hermitian Hamiltonian with chiral 
symmetry, a gradient expansion identifies properties of the 
non-Hermitian operator with those of a stochastic Dirac Hamiltonian 
$\hat{\cal H}_D$. This correspondence allows for the explicit construction of 
eigenfunctions of the non-Hermitian Hamiltonian for individual realizations 
of the disorder. In the thermodynamic limit, 
the wavefunctions were shown to be delocalized
along {\it both} the directions parallel {\it and} perpendicular to ${\bf h}$.

Finally we note that the main characteristics of the 
wavefunction~(\ref{general solution}), long-ranged disorder dependence 
encoded in the `gauge fields', and exponential amplification of the 
wavefunction modulus, can manifest themselves in more complex phenomena than 
that discussed in this Letter: Firstly, one can envisage systems with 
non-trivial topology and/or stochastic potentials with superimposed regular 
structures (e.g. spatially non-uniform distribution functions). In such cases 
the behavior of the wavefunctions may change {\it qualitatively} (e.g. to
localization). 
Secondly, we note that the sensitivity of the
modulus to disorder results in {\em strong statistical fluctuations} of 
the wavefunctions which are large in comparison to the average. In 
particular, correlation functions of the moments, $C$ acquire scaling 
exponents with non-linear dependence in $q_i$, a characteristic related to 
multi-fractality.

We are indebted to A. Andreev, P.W. Brouwer, J.T. Chalker, D.R. 
Nelson, D. Taras-Semchuk and X.-G. Wen for useful discussions.
CM acknowledges a fellowship from the Swiss Nationalfonds.

\end{multicols}

\newpage


\begin{references}
\bibitem{foot4} The components $\chi_\|$ and $\chi_\bot$ are
obtained by solving the two-dimensional Poisson equation in the presence
of the source $\partial_\mu A_\mu$ and 
$\epsilon_{\mu \nu} \partial_\mu A_\nu$, respectively.
The solution $\chi_\|$ ($\chi_\bot$) is unique up to some constant
provided we assume that the support of
$\partial_\mu A_\mu$ ($\epsilon_{\mu \nu} \partial_\mu A_\nu$)
is compact.
\bibitem{foottop} It is interesting to note that, for a system of
non-trivial topology, ${\rm Im}\ z$ induces a twist in the boundary condition
(analogous to an Aharanov-Bohm flux) and can not, in general, be absorbed by 
a gauge transformation.
\bibitem{foot7} 
The thermodynamic limit is taken after imposing periodic boundary conditions
in the finite system.
\end{references}
\end{document}